\input{aipcheck}

\documentclass[
    ,final            % use final for the camera ready runs
  ]
  {aipproc}

\layoutstyle{6x9}
\usepackage{amssymb}

\begin{document}

	\title{Probing gluon nuclear PDF with direct photon production in association with a heavy quark}


	\author{K.~Kova\v{r}\'{\i}k}{
	  address={Institute for Theoretical Physics, Karlsruhe Institute of Technology, Karlsruhe, D-76128,Germany}
	}
	\author{T.~Stavreva}{
	  address={Laboratoire de Physique Subatomique et de cosmologie, Grenoble, F-38026, France}
	}

\begin{abstract}
	We investigate a possible use of direct photon production in
	association with a heavy quark in $pA$ collisions at the large hadron collider (LHC) to 
	constrain the nuclear gluon parton distribution function (PDF).
	This process is sensitive to both, the nuclear heavy quark and gluon parton
	distribution functions and is a very promising candidate to help determine the gluon nuclear PDF
	which is still largely untested.
\end{abstract}

\maketitle

%%%%%%%%%%%%%%%%%%%%%%%%%%%%%%%%%%%%%%%%%%%%
%% MAINMATTER
%%%%%%%%%%%%%%%%%%%%%%%%%%%%%%%%%%%%%%%%%%%%
%
\section{Introduction}
Parton distribution functions (PDFs) are an essential component of any prediction involving colliding hadrons. 
The PDFs are non-perturbative objects which have to be determined from experimental input and link theoretical 
perturbative QCD predictions to observable phenomena at hadron colliders. 
In view of their importance, the proton PDFs have been a focus of long and dedicated global analyses performed 
by various groups.
Over the last decade, global analyses of PDFs in nuclei --~or nuclear PDFs (nPDFs)~-- have been performed by several groups:
nCTEQ~\cite{Schienbein:2009kk,Kovarik:2010uv}, nDS~\cite{deFlorian:2003qf}, EPS09~\cite{Eskola:2009uj},  and  HKN07~\cite{Hirai:2007sx}.
In a manner analogous to the proton PDFs, the nPDFs are needed in order to predict observables in
proton--nucleus ($pA$) and nucleus--nucleus ($AA$) collisions. 
However, as compared to the proton case, the nuclear parton distribution functions are far less well constrained.
Data that can be used in a global analysis are available for fewer hard processes and also cover a smaller 
kinematic range.
Here, we focus on constraining parton densities in nuclei using the production of a direct photon in association with a 
heavy-quark jet in $pA$ collisions (all details of the analysis presented here can be found in \cite{Stavreva:2010mw}).
\section{Direct photons}
Single direct photons have long been considered an excellent probe of the structure of the proton due to 
their point-like electromagnetic coupling to quarks and due to the fact that they escape 
confinement.  
Their study can naturally be extended to high-energy nuclear collisions where one can use 
direct photons to investigate the structure of nuclei as well. 
Concentrating on a double inclusive production of a direct photon with a heavy quark allows access to different PDF components.
Single direct photons couple mostly to valence quarks in the proton or nuclei. By investigating direct photons accompanied by
heavy quark jets, one gains access to the gluon and the heavy quark PDF. That is because, at leading order ${\cal O}(\alpha\alpha_s)$, the direct 
photon with a heavy quark arises only from $g Q \to \gamma Q$ Compton scattering process as opposed to the single photon in which case
a Compton scattering contribution $gq \to \gamma q$ competes against a contribution from quark annihilation $q \bar q \rightarrow \gamma g$.
At leading order, we see that the initial state for the direct photon production with a heavy quark jet depends only on gluon PDF and heavy 
quark PDF where also the latter is often radiatively generated from the gluon leading to even stronger dependence of this process 
on the gluon PDF.
\section{Probing nuclear gluon PDF}
In order to obtain results in hadronic collisions, the partonic NLO cross-section for direct photon in association with
a heavy quark has to be convoluted with PDFs for protons and/or nuclei \cite{Stavreva:2009vi}.
For nuclei we show results using the most recent nCTEQ~\cite{Schienbein:2009kk}, EPS09~\cite{Eskola:2009uj}, 
and HKN07~\cite{Hirai:2007sx} nuclear PDF sets.
Each set of nuclear PDFs is connected to a set of proton PDFs to which it reduces in the limit $A \to 1$
where $A$ is the atomic mass number of the nucleus. Therefore we use the various nPDFs together with their corresponding 
proton PDFs in the calculations.
\begin{figure}[t]
\includegraphics[scale=0.26,angle=-90]{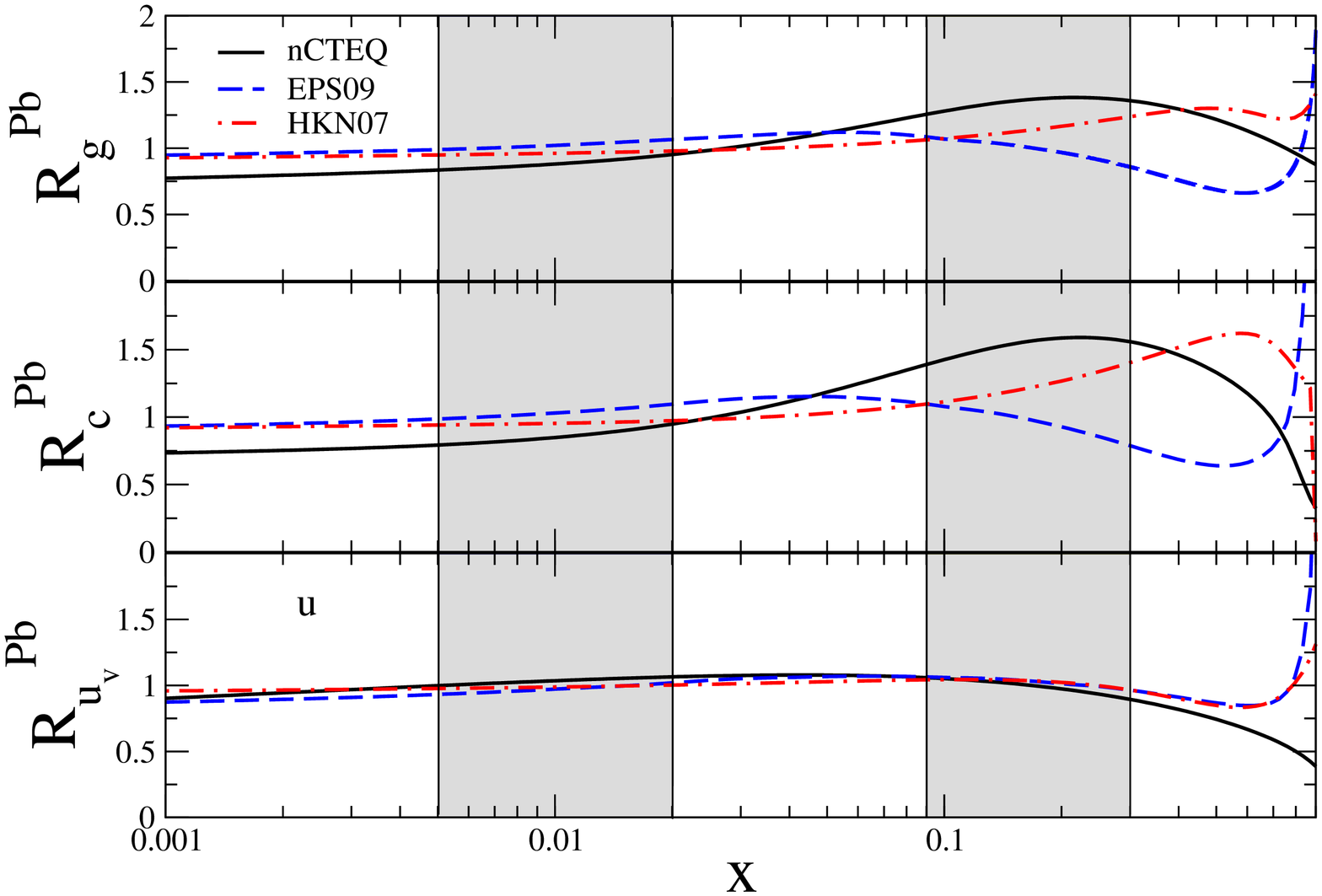}
\includegraphics[scale=0.26,angle=-90]{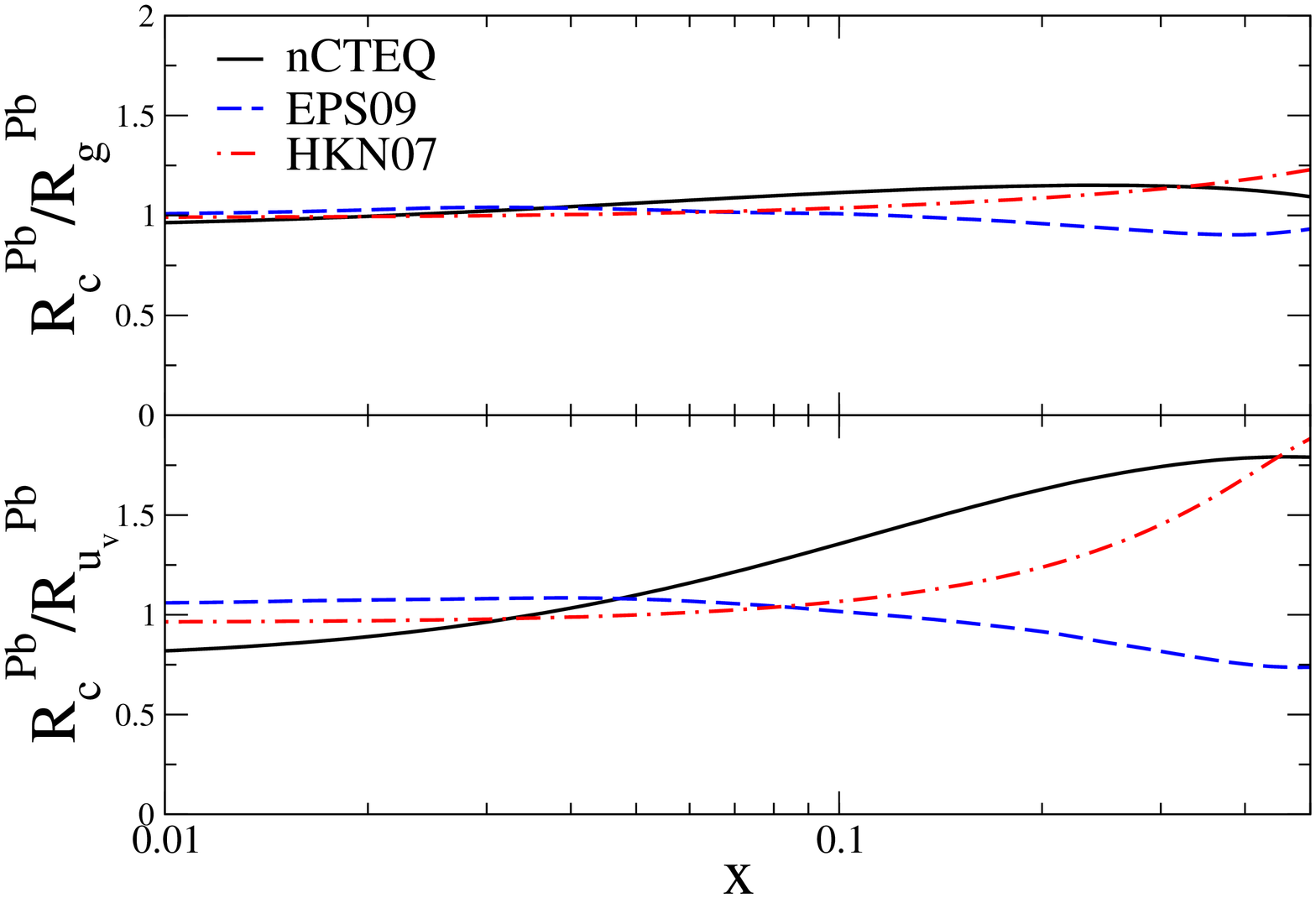}
\caption{Left: nPDF ratios $R_g^{Pb}=g^{p/Pb}(x,Q)/g^{p}(x,Q)$ (top) , 
$R_c^{Pb}=c^{p/Pb}(x,Q)/c^{p}(x,Q)$ (middle), 
$R_{u_v}^{Pb}=u_v^{p/Pb}(x,Q)/u_v^{p}(x,Q)$ (bottom) 
at $Q=50$ GeV within nCTEQ (solid black line), 
EPS09 (dashed blue line), and HKN07 (dash-dotted red line). 
The shaded regions correspond to the $x$-values probed at RHIC ($x\sim 10^{-1}$) and the 
LHC ($x\sim 10^{-2}$). 
Right: double ratios $R_c^{Pb}/R_g^{Pb}$ and $R_c^{Pb}/R_{u_v}^{Pb}$ 
using the same nPDF sets.}
\label{fig-ratio_gcu}
\end{figure}

No current nuclear PDF analysis include a possibility of an intrinsic heavy quark component at the input scale and so the heavy quark 
PDF at a higher scale is generated using the gluon PDF where the gluon splits into heavy quark and anti-quark. Therefore, 
the heavy quark PDF follows in shape the gluon PDF rather closely (see Fig.~\ref{fig-ratio_gcu}) and we will concentrate on discussing
the gluon PDF in more detail.

\begin{figure}
\includegraphics[angle=-90,scale=0.26]{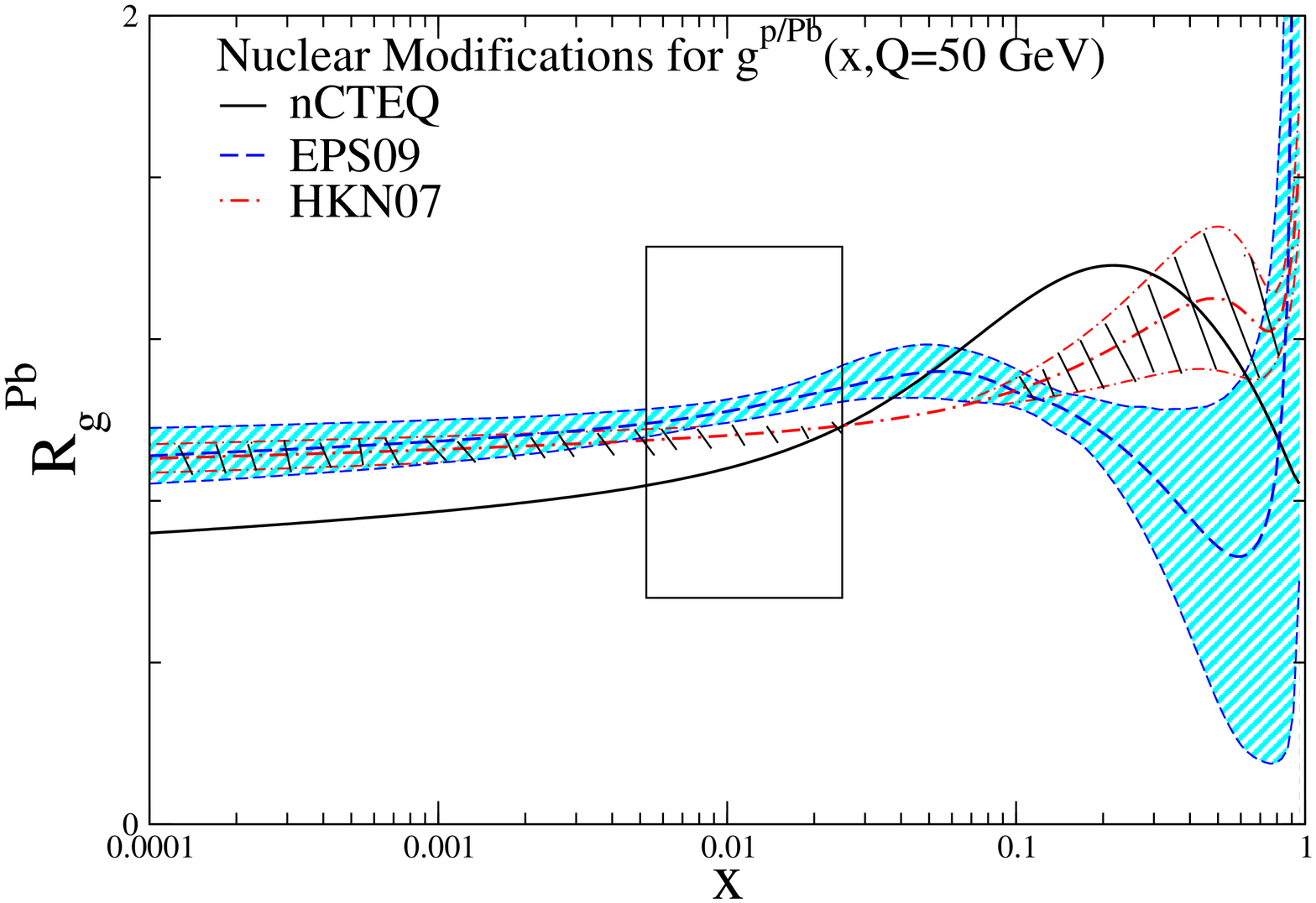}
\includegraphics[angle=-90,scale=0.26]{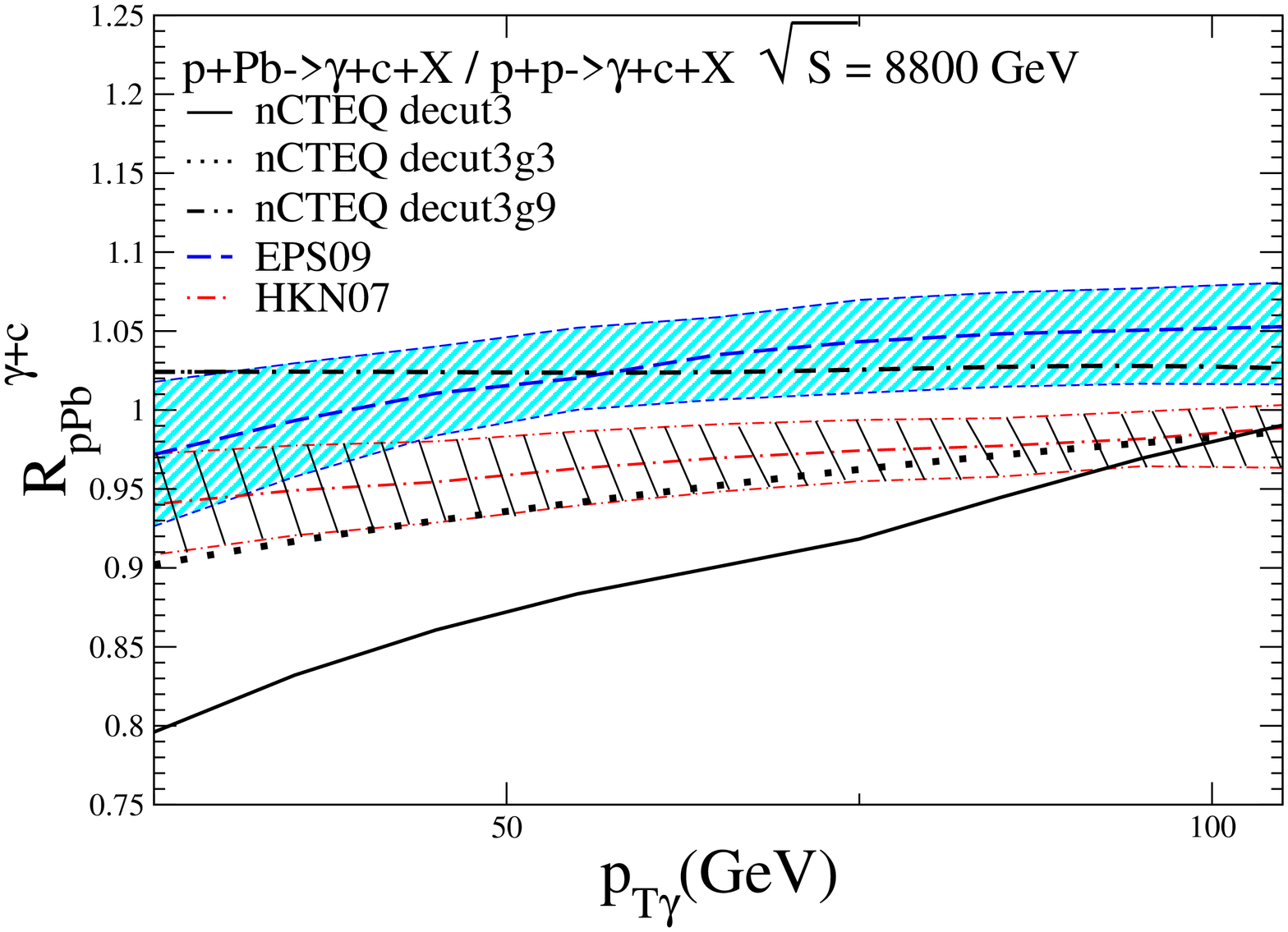}
\caption{Nuclear modifications ${R_g^{A}}=g^{p/A}(x,Q)/g^{p}(x,Q)$.
Left: For lead at $Q=50$ GeV. Shown are results for nCTEQ decut3 (solid, black line),
EPS09 (dashed, blue line) + error band, HKN07 (dash-dotted, red line) + error band.
The box exemplifies the $x$-region probed at the LHC ($\sqrt{s}=8.8$~TeV).
Right: nuclear production ratio of $\gamma+c$ cross-section at LHC within ALICE PHOS acceptances, 
using nCTEQ decut3 (solid black line), nCTEQ decut3g3 (dotted black line), nCTEQ decut3g9 (dash-dot-dashed black line), EPS09 (dashed blue line) + error band, HKN07 (dash-dotted red line) + error band}
\label{fig-gluon}
\end{figure}

A common feature of all nPDF global analyses is that the nuclear gluon distribution is only very weakly
constrained in the $x$-range $0.02\lesssim x \lesssim 0.2$ from the $Q^2$-dependence 
of structure function ratios in deep-inelastic scattering (DIS) $F_2^{Sn}(x,Q^2)/F_2^{C}(x,Q^2)$, 
measured by the NMC collaboration~\cite{Arneodo:1996rv}.
In order to compare the various nPDF sets, we plot in Fig.~\ref{fig-gluon} 	
the gluon distribution ratio $R_g^{A}(x, Q)=g^{p/A}(x, Q)/g^p(x, Q)$ 
as a function of $x$ for a lead nucleus at $Q=50$ GeV.
The chosen hard scale $Q=50$ GeV is typical for direct photon production at
the LHC and the box highlights the $x$-region probed by the LHC.

The fact that the nuclear gluon distribution is poorly constrained is reflected by large PDF uncertainty bands of the gluon
PDF of the HKN07 and EPS09. Moreover, additional uncertainty is connected to the choice of parameterization and other 
assumptions that are an integral part of any PDF analysis. The second source of uncertainties causes uncertainty bands of HKN07 
and EPS09 to not overlap in some regions and causes also the nCTEQ gluon PDF to lie outside of the other uncertainty bands.
Also the rather narrow and overlapping bands at small $x<0.02$ do not reflect any constraints
by data, but instead are theoretical assumptions imposed on the small-$x$ behavior of the gluon distributions.

The nuclear production ratio $R_{pPb}^{\gamma+c}={1\over 208}
{d\sigma/dp_{T\gamma}(p{\rm Pb} \rightarrow \gamma+c+X)
\over d\sigma/dp_{T\gamma}(pp \rightarrow \gamma+c+X)}$
in a kinematic range probed by the ALICE experiment at the LHC
is shown in Fig.~\ref{fig-gluon} (right) using several nuclear PDFs EPS09 (dashed blue line), HKN07 (dash-dotted red line)
and a series of nCTEQ fits \texttt{decut3} (solid black line), \texttt{decut3g3} (dotted black line) and \texttt{decut3g9} 
(dash-dot-dashed black line) which differ in assumptions on the small $x$ behavior of the gluon PDF (details see \cite{Stavreva:2010mw}).
For the first two cases the bands represent the nPDF uncertainties.  
Remarkably, there is almost no overlap between the EPS09, the HKN and nCTEQ predictions, therefore an 
appropriate measurement of this process will be able to distinguish between the nPDF sets.  
\begin{figure}
\includegraphics[angle=-90,scale=0.26]{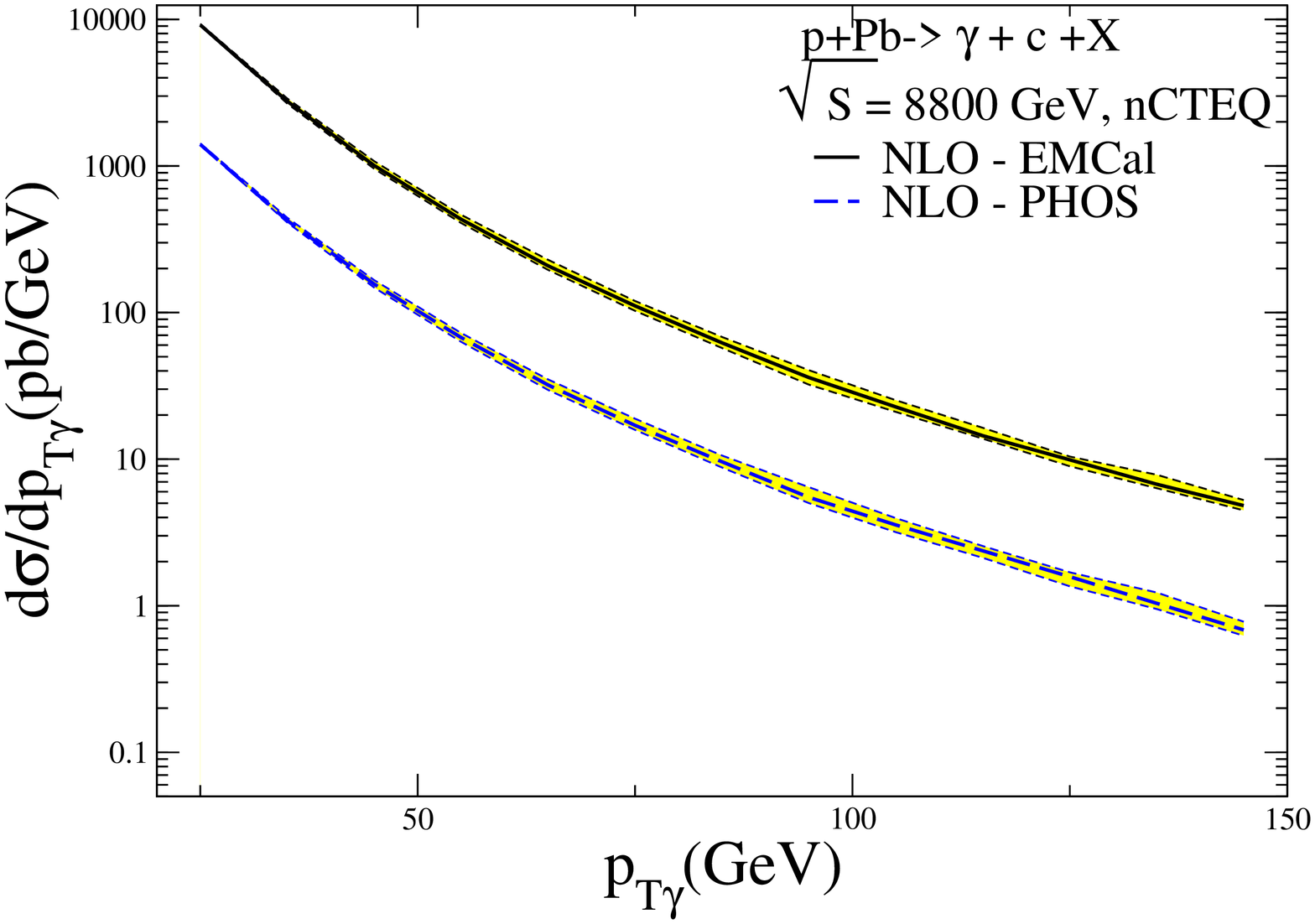}
\includegraphics[angle=-90,scale=0.26]{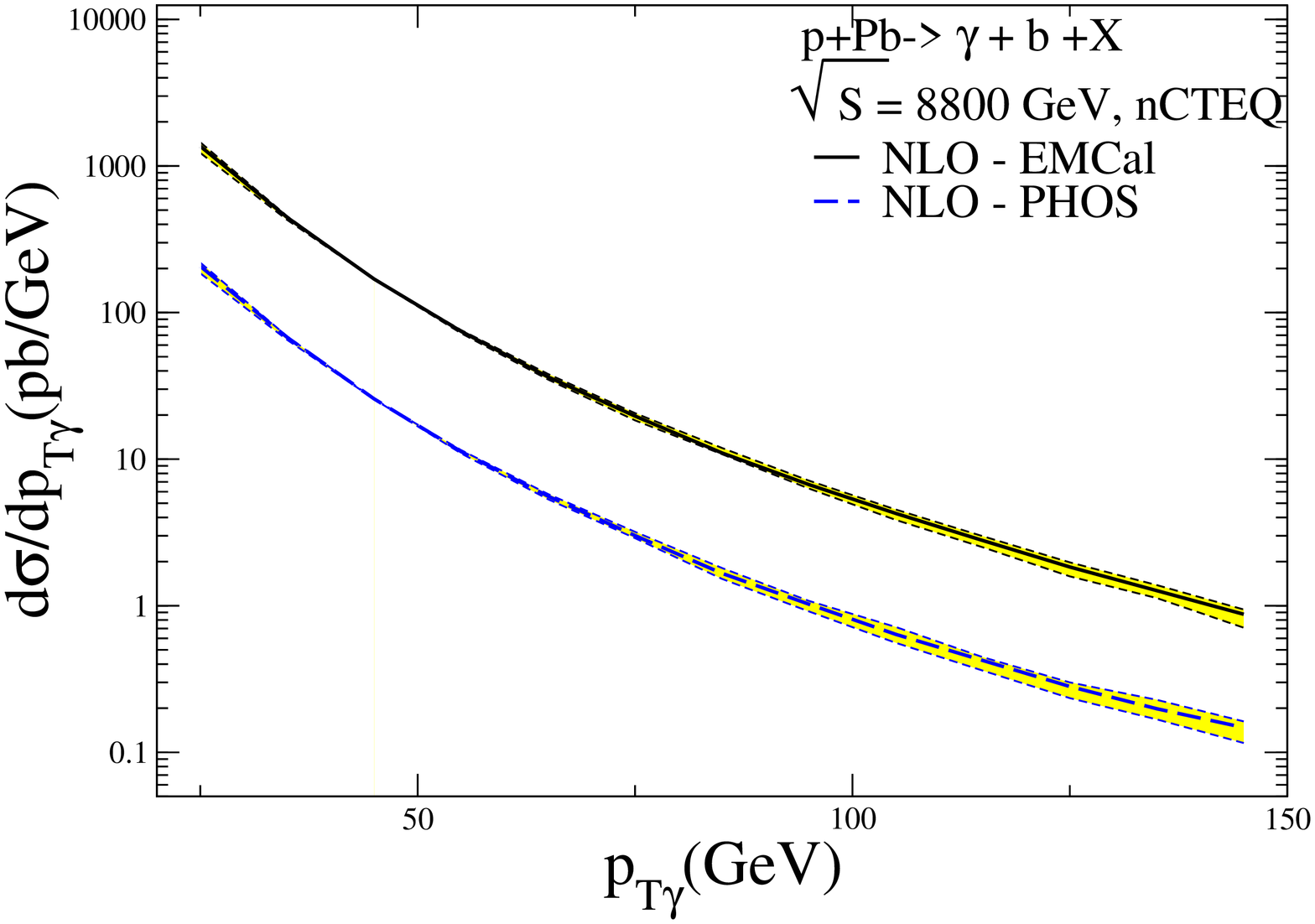}
\caption{NLO differential cross-section for $\gamma+c$ (left) and $\gamma+b$ (right) production in $p$--Pb collisions at a center-of-mass energy of $\sqrt{s}=8.8$ TeV in PHOS (lower band) and EMCal (upper band) acceptances.}
\label{fig-cross-sec2}
\end{figure}

In the ALICE experiment, photons can be identified in the EMCal electromagnetic calorimeter, or in the PHOS 
spectrometer with a somewhat more limited acceptance. In Fig.~\ref{fig-cross-sec2}, we show results of calculations 
carried out for $p$--Pb collisions at the LHC nominal energy $\sqrt{s}=8.8$~TeV using the cuts and acceptances of both 
EMCal and PHOS detectors. The differential NLO cross-section is plotted as a function of the photon transverse momentum 
in the $\gamma+c$ ($\gamma+b$) channel in 
Fig.~\ref{fig-cross-sec2} left (right) for both PHOS (lower band) and EMCal (upper band); 
the dotted curves indicate the theoretical scale uncertainty.

The total integrated cross-section for the EMCal electromagnetic calorimeter is $119000~{\rm pb}$ for $\gamma + c$ process and 
$22700~{\rm pb}$ for $\gamma + b$.
\section{Conclusion}
We have presented a detailed phenomenological study of direct photon production
in association with a heavy-quark jet in $pA$ collisions at the LHC, at next-to-leading order in QCD.
The dominant contribution to this process is given by the 
$gQ\to \gamma Q[+g]$ subprocess. This offers a sensitive mechanism to constrain the heavy-quark and gluon 
distributions in nuclei, whose precise knowledge is necessary in order to predict the rates of hard processes 
in heavy-ion collisions where quark-gluon plasma is expected to be formed.


\bibliographystyle{aipproc}   % if natbib is available

\bibliography{npdf}

%  \typeout{******************************************}
%  \typeout{** Please run "bibtex \jobname" to optain}
%  \typeout{** the bibliography and then re-run LaTeX}
%  \typeout{** twice to fix the references!}
%  \typeout{******************************************}

\end{document}